\documentclass[%
 reprint,
 amsmath,amssymb,
 aps,
]{revtex4-2}

\usepackage{siunitx}
\usepackage{graphicx}
\usepackage{dcolumn}
\usepackage{bm}

\begin{document}

\preprint{APS/123-QED}

\title{Decoupling elasticity and electrical conductivity of carbon black gels filled with insulating non-Brownian grains}

\author{Thomas Larsen}
 \altaffiliation[Also at ]{Advent Technologies A/S, Lyngvej 8, 9000 Aalborg, Denmark.}
 \email{larsenn.thomas@gmail.com}
\affiliation{%
 Department of Materials and Production, Aalborg University\\
 Fibigerstr\ae{}de 16, 9220 Aalborg \O{}st, Denmark
}%

\author{John R. Royer}
\author{Fraser H. J. Laidlaw}
\author{Wilson C. K. Poon}
\affiliation{
 SUPA, School of Physics and Astronomy, The University of Edinburgh \\
 King's Buildings, Peter Guthrie Tait Road, Edinburgh EH9 3FD, United Kingdom
}%

\author{Tom Larsen}
\author{S\o{}ren J. Andreasen}
\affiliation{%
 Advent Technologies A/S\\
 Lyngvej 8, 9000 Aalborg, Denmark
}%

\author{Jesper de C. Christiansen}
\affiliation{%
 Department of Materials and Production, Aalborg University\\
 Fibigerstr\ae{}de 16, 9220 Aalborg \O{}st, Denmark
}%


\begin{abstract}
A unique bistable transition has been identified in granular/colloidal gel-composites, resulting from shear-induced phase separation of the gel phase into dense blobs. In energy applications, it is critical to understand how this transition influences electrical performance. Mixing conductive colloids with conductive inclusions, we find the conductivity and elasticity move in concert, both decreasing in the collapsed phase-separated state. Surprisingly, with insulating inclusions these properties can become decoupled, with the conductivity instead increasing despite the collapse of the gel structure.
\end{abstract}

\maketitle


\section{Introduction}

Attractive colloidal particles are subject to Brownian motion which leads to the formation of a space-spanning network with a yield stress. These colloidal gels are used across various industries, from cosmetic and personal care products \cite{article:Joshi2014} to energy materials \cite{article:Spahr2011,article:Khandavalli2018,article:Wei2015}. The microstructure and rheology of gels are closely connected \cite{article:Koumakis2015}, allowing control over these properties by means of shear deformation \cite{article:Moghimi2017}. Thus, a thorough understanding of their rheology is vital. 

Carbon black (CB) gels are well studied \cite{article:Richards2023} and exhibit intriguing properties such as strong flow-history dependence of their microstructure \cite{article:Osuji2008,article:Varga2019}, mechanical \cite{article:Ovarlez2013,article:Wang2022,article:Hipp2019,article:Dages2022} and electrical \cite{article:Helal2016,article:Narayanan2017,article:Youssry2013} properties. Adding polymers to CB gels modifies their rheo-electric behavior \cite{article:Legrand2023}, while the impact of solid filler particles on the composite rheology is non-trivial \cite{article:King2008} and conflicting with the assumption of an additive effect of each constituent which is otherwise not uncommon to assume \cite{article:Ma2019}. 

Recently, the mixing of gels with large non-Brownian fillers has led to the discovery of a bistable regime \cite{article:Jiang2022} where flow acts as a `switch' between solid-like and liquid-like states. Such a transition was subsequently demonstrated also in oil-based CB/graphite mixtures \cite{article:Larsen2023_rheolacta} where a low-yield-stress (``liquid-like") composite exhibited poorer conductivity compared to a high-yield-stress (``solid-like") one. The bistability-phenomenon appears generic to such binary composites, yet the contribution of the individual constituents to the total composite properties remains elusive \cite{article:Jiang2023,article:Larsen2023_rheolacta,article:Jiang2023_2,article:Li2023}.

In this work, we primarily study the influence of insulating non-Brownian hollow glass spheres (HGSs) on the rheo-electric properties of a mineral oil-based CB gel. Similar to graphite \cite{article:Larsen2023}, the HGSs form an adhesive suspension when dispersed in mineral oil \cite{article:Papadopoulou2020,SM}. However, the current model system rectifies a shortcoming of these previous studies \cite{article:Larsen2023_rheolacta}: it allows us to isolate the effect of conductivity changes to the gel.

\section{Materials and methods}

\subsection{Materials}

Heavy mineral oil (Merck 330760, viscosity \SI{0.20}{\pascal\second} at \SI{20}{\celsius}) was used as the suspending medium. Carbon black (Ensaco 250G) with an oil adsorption number (OAN) of \SI{190}{mL/100 \ g} (measured by the manufacturer) and synthetic graphite (Timrex KS150) was kindly provided by Imerys Graphite \& Carbon (Bodio, Switzerland). Hollow glass spheres (Sigma-Aldrich 440345) with a mean particle size of $9-13 \ \si{\mu m}$ were chosen as the insulating large fillers.

To prepare the composites, the CB and oil was vortex mixed before bath sonicating the mixture for at least \SI{90}{min} to achieve a visibly homogeneous mixture. The resulting gel was left to rest for 24 hours. To prepare the composite, HGSs were vacuum dried at \SI{60}{\celsius} for a minimum of five hours prior to vortex mixing with the CB gel. The resulting binary composite was left to rest overnight before any testing. The final samples were stored at room temperature in a desiccator. Before loading into the rheometer, the samples were vortex mixed and stirred vigorously. Graphite-based composites were prepared and treated similarly.

To obtain the dry volume fraction of CB $\phi_{\rm CB,dry}$ we used $\rho_{\rm CB}=1.9 \ \si{g \cdot cm^{-3}}$ as the density. The OAN was used to account for the porosity of CB aggregates, with more porous structures possessing a larger OAN \cite{article:Richards2023}. Following refs. \cite{article:Liu2023,article:Lin2007}, we used the OAN to estimate the maximum volume fraction of CB in the suspending liquid $\Phi_{\rm CB,pagg} = 0.22$. Then, the effective CB volume fraction was calculated as $\phi'_{\rm eff,CB} = \phi_{\rm CB,dry}/\Phi_{\rm CB,pagg}$ \cite{article:Richards2017}. Due to the volume occupied by the large fillers in the binary composites, we determined a modified effective CB volume fraction by subtracting the large filler volume fraction $\phi_{\rm f}$ from the total composite volume fraction  

\begin{equation}
    \phi_{\rm eff,CB} = \frac{\phi'_{\rm eff,CB}}{1-\phi_{\rm f}}.
\end{equation}

We calculated $\phi_{\rm f}$ based on the densities $\rho_{\rm HGS}=1.1 \ \si{g \cdot cm^{-3}}$ and $\rho_{\rm Graphite}=2.22 \ \si{g \cdot cm^{-3}}$ of glass spheres and graphite, respectively, measured by the manufacturers.

\subsection{Rheometry}

Steady shear rheometry was performed on a stress-controlled HR-20 rheometer (TA Instruments) with the temperature kept at \SI{20}{\celsius} using a Peltier plate. A crosshatched parallel plate geometry (diameter \SI{40}{\milli\meter}, gap height \SI{800}{\micro\meter}) was used to minimize wall slip. To erase memory of the loading history, the CB gel and binary composites were sheared at $\dot{\gamma} = 500 \ \si{s^{-1}}$ for \SI{600}{s} followed by \SI{600}{s} rest. Sample edge fracture was absent at all probed shear rates as confirmed by visual inspection. Shear rates are reported at the rim.

Flow curves were obtained by rejuvenating the sample at $\dot{\gamma} = 500 \ \si{s^{-1}}$ for \SI{300}{s} after which the shear rate was either ramped down or stepped down in discrete steps to the lowest shear rate probed, $\dot{\gamma} = 0.01 \ \si{s^{-1}}$. The ramp-down protocol consisted in progressively lowering the shear rate using a fixed time at each $\dot{\gamma}$, and the last data point was used for analysis. The step-down protocol consisted in rejuvenating ($\dot{\gamma} = 500 \ \si{s^{-1}}$ for \SI{300}{s}) the sample before stepping directly down to a lower $\dot{\gamma}$. At low $\dot{\gamma}$, the transient stress response exhibited an overshoot in both the gel and composites, and so, for the step-down protocol we used the maximum transient $\tau$ for data analysis.

Oscillatory shear strain amplitude sweeps were performed to determine the moduli in the linear viscoelastic range and the yield strain (where $G'=G''$). During the strain amplitude up-sweep (starting at $\gamma_0=0.0001$) the angular frequency was $\omega=10 \ \si{rad \cdot s^{-1}}$. Each data point was acquired by first conditioning the sample with eight oscillation cycles followed by eight cycles over which data was averaged.

\subsection{Dielectric measurements}

A TRIOS-controlled Keysight E4980A LCR meter was used with the HR-20 rheometer for the dielectric spectroscopy measurements which were all performed on samples at rest. Smooth parallel plates (diameter \SI{25}{\milli\meter}, gap height \SI{800}{\micro\meter}) were used as electrodes. Short- and open-circuit measurements were conducted to correct the measured data, obtained by performing logarithmically spaced frequency sweeps from \SI{20}{Hz} to \SI{1}{MHz} at a voltage amplitude of \SI{100}{mV} which ensured an Ohmic behavior of the samples. Before collecting the dielectric data, samples were allowed to rest for \SI{600}{s} following their deformation.

The AC conductivity was calculated as $\sigma_{\rm ac} = \omega \epsilon_0 \epsilon_{\rm r}''$ where $\epsilon_0 \simeq 8.854\cdot 10^{-12} \ \si{F \cdot m^{-1}}$ is the permittivity of free space, $\omega=2\pi f$ is the angular frequency of the applied field, and $\epsilon_r''$ the imaginary part of the complex permittivity $\mathrm{Im}\left(\epsilon^*_r \right) = \epsilon_{\rm r}''$ \cite{book:Schonhals2003}. On its own, the heavy mineral oil exhibited a storage permittivity $\epsilon_{\rm r}' \simeq 2$ and negligible conductivity. AC conductivities are reported at $f=126.2 \ \si{Hz}$ and referred to as the low-frequency conductivity $\sigma$; this was the lowest frequency at which reproducible data could be collected on all samples.

\subsection{Cryogenic scanning electron microscopy}

The scanning electron microscope (SEM) was a Zeiss Crossbeam 550 FIB-SEM with a Quorum Technologies Ltd PP3010T cryogenic attachment. After shearing the samples on the rheometer, the upper geometry was lifted and, carefully, a small amount of sample was transferred and placed between two rivets. The rivets were plunge frozen into slush nitrogen before being loaded into the Quorum chamber. The samples were freeze fractured inside the Quorum chamber at \SI{-140}{\celsius}. The uncoated samples were imaged with a \SI{1}{kV} acceleration voltage, acquiring micrographs using both the Zeiss 'InLens' secondary electron detector, located in the electron column, along with the secondary electron secondary ion (SESI) detector to differentiate topographic features and compositional changes.

\section{Results and discussion}

\begin{figure}
    \centering
    \includegraphics{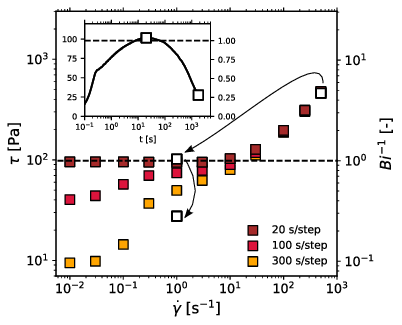}
    \caption{Flow curves, obtained by ramping down the shear rate $\dot{\gamma}$ of a composite with $\phi_{\rm HGS}=0.20$ and $\phi_{\rm eff,CB}=0.17$ in heavy mineral oil. The horizontal dashed line indicates the yield stress of a rejuvenated composite $\tau^{\rm H}_{\rm y} = 98 \ \si{Pa}$. It is used to define the inverse Bingham number ${\it Bi}^{-1}=\tau/\tau^{\rm H}_{\rm y}$. Here $\tau$ is the measured shear stress. The open squares show the maximum and minimum stress measured over \SI{1800}{s} at $\dot{\gamma}=1 \ \si{s^{-1}}$. Inset: transient stress-response after stepping down to $\dot{\gamma}=1 \ \si{s^{-1}}$ with the open squares showing the values plotted in the main figure. The horizontal line indicates $\tau^{\rm H}_{\rm y}$.}
    \label{fig:1}
\end{figure}

The salient rheological features of a composite with volume fractions $\phi_{\rm HGS}=0.20$ and $\phi_{\rm eff,CB}=0.17$ of HGSs and CB, respectively, are demonstrated in Figure \ref{fig:1}. A series of shear rate ramp-downs at increasingly slower rates result in weaker composites with the resulting flow curves departing below the yield stress of the homogeneous state $\tau^{\rm H}_{\rm y} = 98 \ \si{Pa}$; a characteristic of bistable composites \cite{article:Jiang2022,article:Larsen2023_rheolacta}. To demonstrate that the rheology becomes heavily strain-dependent below $\tau^{\rm H}_{\rm y}$ (or, equivalently, when the inverse Bingham number \cite{article:Hipp2019} ${\it Bi}^{-1}=\tau/\tau^{\rm H}_{\rm y}\approx 1$, with $\tau$ being the shear stress) we quench the composite from a high-shear state to a shear rate $\dot{\gamma}=1 \ \si{s^{-1}}$ and accumulate up to $\gamma_{\rm acc}=\dot{\gamma}t=1800$ strain units. The transient response (Figure \ref{fig:1} inset) shows a strong recovery-then-decay response similar to oil-based CB gels \cite{article:Wang2022} where it is associated with a transition from small fractal agglomerates to large non-fractal ones \cite{article:Hipp2019}.

\begin{figure}
    \centering
    \includegraphics{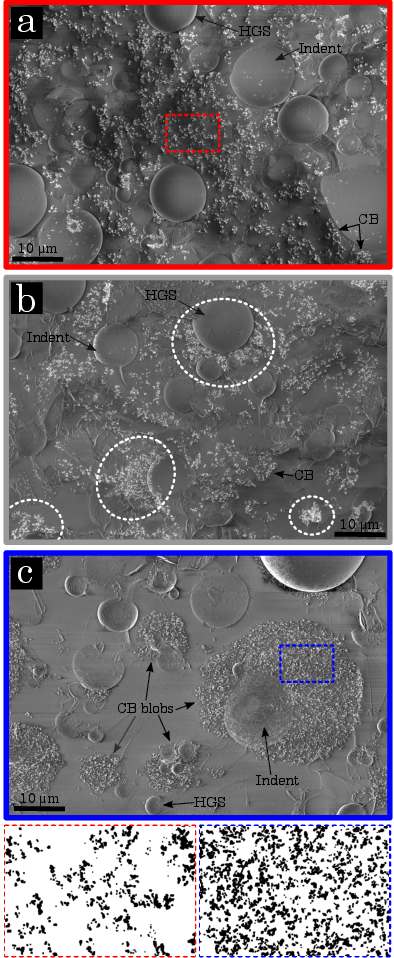}
    \caption{Scanning electron microscopy images of composites with $\phi_{\rm HGS}=0.20$ and $\phi_{\rm eff,CB}=0.17$ after (a) rejuvenation and steady shear at $\dot{\gamma}=1 \ \si{s^{-1}}$ for (b) \SI{300}{s} and (c) \SI{1800}{s}. The highlighted regions in (b) show developing blobs of carbon black (CB). The boxed regions in (a,c) have been binarized for image analysis with the results shown at the bottom (see \cite{SM} for additional images). The area fraction of CB in (a) was found to be $\sim 0.16 \pm 0.02$ while in (c) it was $\sim 0.30 \pm 0.02$ specifically inside blobs.}
    \label{fig:2}
\end{figure}

We therefore use cryogenic scanning electron microscopy to image a composite with $\phi_{\rm HGS}=0.20$ and $\phi_{\rm eff,CB}=0.17$ at various stages during shear at $\dot{\gamma}=1 \ \si{s^{-1}}$. Rejuvenation results in a homogeneous microstructure (Figure \ref{fig:2}a), whereas accumulating $\gamma_{\rm acc}=300$ leads to incipient phase separation of CB into blobs (Figure \ref{fig:2}b). Reaching $\gamma_{\rm acc}=1800$ produces a severely phase separated microstructure (Figure \ref{fig:2}c). The microstructural transition produced by the ramp-down protocol, Figure \ref{fig:1}, is analogous~\cite{SM}. We find that blobs in the phase separated state have diameters $d_{\rm blob}=16 \pm 4 \ \si{\mu m}$ and $20 \pm 11 \ \si{\mu m}$ in composites with $\phi_{\rm HGS}=0.20$ and  either $\phi_{\rm eff,CB}=0.056$ or $\phi_{\rm eff,CB}=0.17$, respectively. These sizes are remarkably consistent with the observed blob diameters in the silica-based binary composites \cite{article:Jiang2022} where $d_{\rm blob}\approx 20 \ \si{\mu m}$. Similar to the gel-phase blobs in refs. \cite{article:Jiang2022,article:Larsen2023_rheolacta}, they appear notably denser compared to the gel in rejuvenated composites which is confirmed by image analysis; the local area density of CB in blobs is $\approx30 \ \%$ in composites with $\phi_{\rm eff,CB}=0.056$ and $\phi_{\rm eff,CB}=0.17$~\cite{SM}. 

\begin{figure*}
    \centering
    \includegraphics{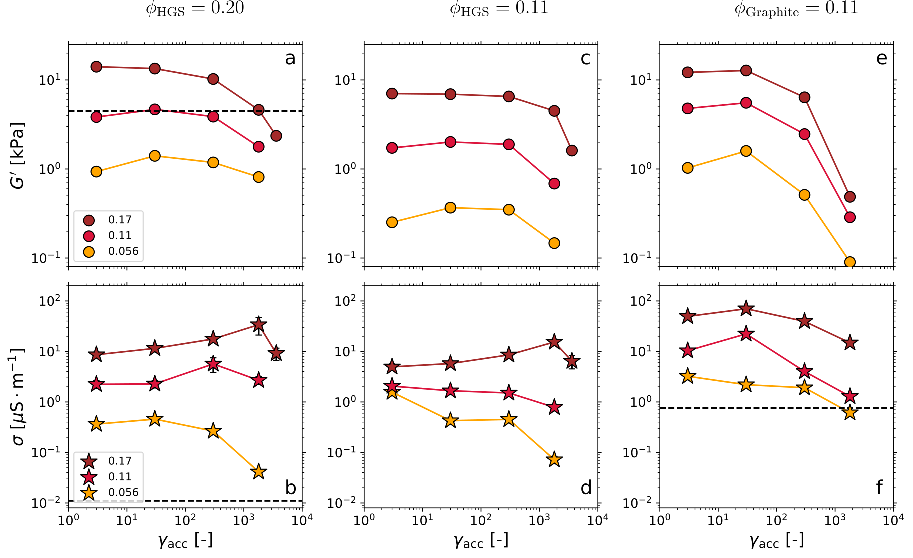}
    \caption{Storage modulus $G'$ and low-frequency electrical conductivity $\sigma$ for composites with effective carbon black volume fractions $\phi_{\rm eff,CB}$ in the legend and glass sphere volume fractions (a,b) $\phi_{\rm HGS}=0.20$ and (c,d) $\phi_{\rm HGS}=0.11$, and graphite volume fraction (e,f) $\phi_{\rm Graphite}=0.11$. All composites were exposed to the same shear history, specifically by \textit{pre}-shearing at $\dot{\gamma}=500 \ \si{s^{-1}}$ (rejuvenation) before stepping down to $\dot{\gamma}=1 \ \si{s^{-1}}$ for a time $t$ to accumulate $\gamma_{\rm acc}=\dot{\gamma}t$ strain units. The dashed horizontal lines show data from rejuvenated pure large filler-suspensions ($\phi_{\rm eff,CB}=0$).}
    \label{fig:3}
\end{figure*}

We expect the microstructural transition to be reflected in the rheo-electric properties of the composites, so following an accumulated strain $\gamma_{\rm acc}$ at $\dot{\gamma}=1 \ \si{s^{-1}}$ the storage modulus $G'$ in the linear viscoelastic region and low-frequency conductivity $\sigma$ are measured. In composites with $\phi_{\rm HGS}=0.20$, a notable weakening of the storage modulus results from the accumulation of strain (Figure \ref{fig:3}a) as may be expected from the severely phase separated microstructure. At $\phi_{\rm eff,CB}=0.056$, $G'$ is almost constant but of lower magnitude than the pure HGS suspension ($G'\simeq 4.5 \ \si{kPa}$) which may be related to gel disturbance by large fillers as reported in ref. \cite{article:Jiang2023}, yet the exact mechanism is currently unclear.

Surprisingly, when $\phi_{\rm eff,CB}\geq0.11$ a significant increase in $\sigma$, and thus delayed weakening compared with $G'$, is observed (Figure \ref{fig:3}b). Such decoupling is highly concentration-dependent and is a general feature of our HGS-based composites since the same aspects recur when $\phi_{\rm HGS}$ is lowered to \SI{0.11}{} (Figure \ref{fig:3}c,d). Previous work either assumed the mechanical and electrical network properties in binary composites would follow each other \cite{article:Jiang2022} or revealed their mutual coupling in a narrow parameter space \cite{article:Larsen2023_rheolacta,article:Sullivan2022}. Although we demonstrate for graphite-based composites that both $G'$ and $\sigma$ exhibit a modest maximum at $\gamma_{\rm acc}=30$ before subsiding (Figure \ref{fig:3}e,f), indicating their mutual coupling, such behavior is clearly not universal.

\begin{figure}
    \centering
    \includegraphics[width=.8\columnwidth]{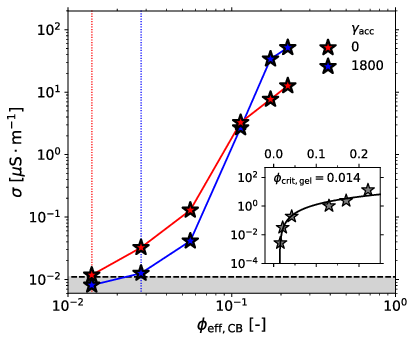}
    \caption{Low-frequency electrical conductivity $\sigma$ for composites with various effective carbon black volume fractions $\phi_{\rm eff,CB}$ and glass sphere volume fraction $\phi_{\rm HGS}=0.20$. Conductivities were measured at rest \SI{1000}{s} after rejuvenating and shearing at $\dot{\gamma}=1 \ \si{s^{-1}}$ for $t=$ \SI{0}{s} or \SI{1800}{s} $\left(\gamma_{\rm acc} = \dot{\gamma} t \right)$. The dotted vertical lines are estimates of the gel-phase electrical percolation thresholds while the horizontal dashed line at $\sigma\simeq1.1\times 10^{-2} \ \mu\si{S\cdot m^{-1}}$ is the conductivity of a pure $\phi_{\rm HGS}=0.20$ suspension. Inset: $\sigma$ of pure gels as a function of $\phi_{\rm eff,CB}$ with a power law fit (solid line) predicting an electrical percolation threshold $\phi_{\rm crit,gel}=0.014$. Accumulated strain in the pure gel was $\gamma_{\rm acc}=0$ after pre-shearing.}
    \label{fig:5}
\end{figure}

\begin{figure*}
    \centering
    \includegraphics{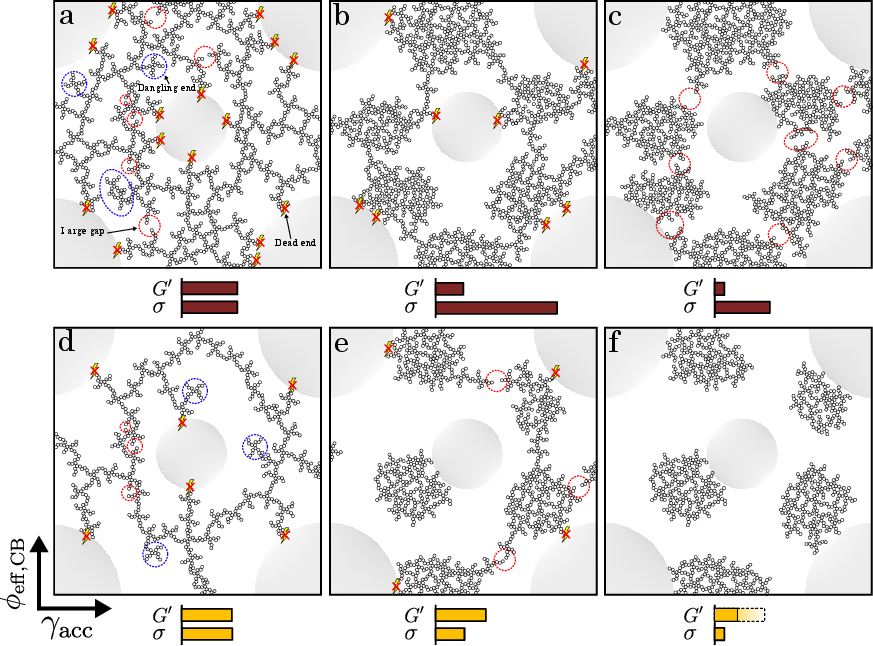}
    \caption{\textbf{Top}: Illustration of the microstructural evolution of a composite with high content ($\phi_{\rm eff,CB}$) of conductive carbon black (black aggregates). After rejuvenation (a), dangling ends (blue circles), dead ends (crossed-over lightning bolt) between gel and glass spheres (filled grey circles), and large gaps (red circles) between carbon black (CB) aggregates limit the conductivity $\sigma$. However, the dead ends provide load-transfer pathways beneficial to the elasticity $G'$. As the accumulated strain $\gamma_{\rm acc}$ increases under low-intensity shear, the colloidal gel phase separates into sparsely-connected dense blobs (b). The blobs provide little rigidity, causing a decrease in $G'$. However, their dense nature leads to short hopping distances between \textit{intra}-blob CB aggregates which, in combination with electrical percolation, increases $\sigma$. Upon further strain accumulation (c), the \textit{inter}-blob gaps eventually become large enough that $\sigma$ decreases along with $G'$. \textbf{Bottom}: At low $\phi_{\rm eff,CB}$, the gel and glass spheres are well dispersed after rejuvenation (d). Increasing $\gamma_{\rm acc}$ under low-intensity shear leads to phase separation of the gel into blobs (e) but with fewer \textit{inter}-blob connections, causing a decrease in $\sigma$. Upon further increasing $\gamma_{\rm acc}$, $\sigma$ decays due to continued phase separation (f). The effect on $G'$ depends on $\phi_{\rm HGS}$ since a high content would render $G'$ relatively unchanged (dashed lines) whereas a low $\phi_{\rm HGS}$ would weaken the composite (full lines).}
    \label{fig:4}
\end{figure*}

Our results reveal that at large $\phi_{\rm eff,CB}$ in the binary HGS-based composites, conductivity benefits from phase separation only up to a point before subsiding. Simulations \cite{article:Wang1998,article:Bao2012,article:Tarlton2017} and experiments on networks of carbon nanotubes \cite{article:Alig2012,article:Gnanasekaran2014,article:Gnanasekaran2016,article:Aguilar2010} and CB \cite{article:Kim2008,article:Zhang2019} have shown that agglomeration of the conductive phase into dense electrically connected clusters can be beneficial to the composite conductivity due to the formation of intimate contacts between the conductors \cite{article:Barsan2016}. Considering that the local CB concentration in blobs is substantially above the bulk concentration (Figure \ref{fig:2}a,c), the resulting intimate CB-CB contacts may increase the total composite conductivity provided an electrically percolated network exists.

A consequence of the localization of CB in dense blobs should be a modified electrical percolation threshold. For $\phi_{\rm eff,CB} \gtrsim 0.11$, rejuvenating the composite produces inferior conductivities compared to microstructures after $\gamma_{\rm acc}=1800$ (Figure \ref{fig:5}). However, at lower $\phi_{\rm eff,CB}$ the rejuvenated microstructure yields higher conductivities due to the large separation between CB blobs at low $\phi_{\rm eff,CB}$. Thus, blob formation leads to a larger electrical percolation threshold ($\approx 0.03$) compared to rejuvenation ($\approx 0.01$), the latter producing a threshold comparable to rejuvenated pure gels (Figure \ref{fig:5} inset). Similarly, we find a lower electrical percolation threshold in composites with $\phi_{\rm HGS}=0.11$ and $\phi_{\rm Graphite}=0.11$ after rejuvenation~\cite{SM}.

The above considerations suggest the following microstructural explanation of the conductivity increase-elasticity decrease at large $\phi_{\rm eff,CB}$ in the binary HGS-based composites. Rejuvenation homogeneously disperses the fractal CB aggregates among the large fillers, forming a space-spanning network (Figure \ref{fig:4}a). Some aggregates terminate at the surface of large fillers (dead ends) or form dangling ends while others are separated by relatively large gaps, all of which increase the electrical resistance \cite{article:Steinhauser2016,article:Barsan2016}. Nevertheless, the large fillers serve as additional load-bearing pathways, so that the modulus benefits from dead ends \cite{article:Gravelle2021}. Accumulating several hundred strain units produces dense blobs of CB. String-like bonds between blobs, or small CB agglomerates in the interstitial space between blobs (Section VI in \cite{SM}) may establish electrical connections but provide limited rigidity to the network \cite{article:Park2015,article:Richards2017}. While network rigidity in colloidal suspensions requires a coordination number of 2.4 \cite{article:Valadez-Perez2013}, it is only 2 for electrical percolation, so that some of the bonds that impart elasticity to the network do not contribute to network conductivity \cite{article:Richards2017}. As the (electrically) connected blobs are densely packed, their internal resistance is low \cite{article:Barsan2016,article:Narayanan2017} and the composite conductivity increases (Figure \ref{fig:4}b). Further strain accumulation results in increasingly separated blobs and a concomitant decline in rheo-electric properties (Figure \ref{fig:4}c). Lowering $\phi_{\rm eff,CB}$ produces a tenuous gel network after rejuvenation (Figure \ref{fig:4}d), which upon phase separation consists of mostly disconnected blobs (Figure \ref{fig:4}e). The blobs eventually become isolated (Figure \ref{fig:4}f) causing a continuous decrease in conductivity whereas the impact on $G'$ is dependent on $\phi_{\rm HGS}$; at high $\phi_{\rm HGS}$ the HGS-network still provides some rigidity so that $G'$ remains relatively unchanged, but this becomes unattainable at low $\phi_{\rm HGS}$ so $G'$ diminishes. 

If the insulating glass spheres in Figure \ref{fig:4} are substituted for conductive graphite, the dead ends become electrically `active' and provide pathways for fast charge transport. Therefore, a large density of CB-graphite contacts is beneficial to both $G'$ \textit{and} the conductivity. However, as strain accumulates under low-intensity shear, the joint graphite/CB blobs become sparsely connected \cite{article:Larsen2023_rheolacta} which opposes the conductivity enhancement due to dead ends turning `active'. As a result, both $G'$ and the conductivity are reduced which, contrary to the HGS-based composites, leads to the coupling of these properties. 

\section{Conclusions and outlook}

We have revealed a non-trivial structure-property relationship in binary composites of colloidal gel with non-Brownian fillers. When composites with insulating glass spheres are sheared where ${\it Bi}^{-1}\approx1$, phase separation is induced and the elasticity and conductivity become decoupled. Under identical shearing conditions with the insulating fillers substituted for conductive graphite, the elasticity and conductivity become coupled. At sufficiently high $\phi_{\rm eff,CB}$, the dense blobs formed during phase separation degrade the composite's elasticity but may enhance the gel network conductivity, provided they form an electrically percolated network. A key factor to understanding this decoupling is the different coordination numbers required for network rigidity and electrical percolation. In contrast to previous works \cite{article:Jiang2022,article:Larsen2023_rheolacta} which were largely observational, our findings therefore present a significant advancement in elucidating the physics governing binary composite behavior. 

Beyond establishing means to optimize mixing and processing of industrial slurries, our findings point towards potential applications for these mixtures where writing (microstructure) and reading (electrical conductivity) of memory is required. This, however, requires mapping the flow history-dependence of the microstructure \cite{article:Jamali2020} and a deeper mechanistic understanding of the microstructural transitions.

\section{Data Availability Statement}

The data in this article are available online at \cite{Data}.

\section{Acknowledgements}

The authors wish to thank J. A. Richards for fruitful discussions. Carbon black powder was kindly provided by Imerys Graphite \& Carbon. A grant from the Industrial PhD programme, Innovation Fund Denmark, project 8053-00063B is gratefully acknowledged.

\bibliography{bibliography}

\end{document}